\documentstyle[12pt,epsf]{article}

\input{psfig}

\def\MRSD0{MRSD0$^\prime$ }

\begin{document}
%
\begin{titlepage}
\begin{flushright}
{FERMILAB-PUB-96/020-E} \vspace{-0.01in} \\
{Submitted to PRL Jan. 25, 1996} \\
\end{flushright}
\begin{center}
{\bf Inclusive jet cross section in ${\bar p p}$ collisions at
$\sqrt{s}=1.8$ TeV}
\end{center}
\begin{normalsize}
%
\font\eightit=cmti8
\def\r#1{\ignorespaces $^{#1}$}
\hfilneg
\begin{sloppypar}
\noindent
F.~Abe,\r {14} H.~Akimoto,\r {32}
A.~Akopian,\r {27} M.~G.~Albrow,\r 7 S.~R.~Amendolia,\r {23} 
D.~Amidei,\r {17} J.~Antos,\r {29} C.~Anway-Wiese,\r 4 S.~Aota,\r {32}
G.~Apollinari,\r {27}
% ...
\end{sloppypar}

\vskip .025in
\begin{center}
(CDF Collaboration)
\end{center}

\vskip .025in
\begin{center}
\r 1  {\eightit Argonne National Laboratory, Argonne, Illinois 60439} \\
\r 2  {\eightit Istituto Nazionale di Fisica Nucleare, University of Bologna,
I-40126 Bologna, Italy} \\
\r 3  {\eightit Brandeis University, Waltham, Massachusetts 02254} \\
\r 4  {\eightit University of California at Los Angeles, Los 
Angeles, California  90024} \\  
\r 5  {\eightit University of Chicago, Chicago, Illinois 60637} \\
\r 6  {\eightit Duke University, Durham, North Carolina  27708} \\
\r 7  {\eightit Fermi National Accelerator Laboratory, Batavia, Illinois 
60510} \\
\r 8  {\eightit Laboratori Nazionali di Frascati, Istituto Nazionale di Fisica
               Nucleare, I-00044 Frascati, Italy} \\
\r 9  {\eightit Harvard University, Cambridge, Massachusetts 02138} \\
\r {10} {\eightit Hiroshima University, Higashi-Hiroshima 724, Japan} \\
\r {11} {\eightit University of Illinois, Urbana, Illinois 61801} \\
\r {12} {\eightit Institute of Particle Physics, McGill University, Montreal 
H3A 2T8, and University of Toronto,\\ Toronto M5S 1A7, Canada} \\
\r {13} {\eightit The Johns Hopkins University, Baltimore, Maryland 21218} \\
\r {14} {\eightit National Laboratory for High Energy Physics (KEK), Tsukuba, 
Ibaraki 305, Japan} \\
\r {15} {\eightit Lawrence Berkeley Laboratory, Berkeley, California 94720} \\
\r {16} {\eightit Massachusetts Institute of Technology, Cambridge,
Massachusetts  02139} \\   
\r {17} {\eightit University of Michigan, Ann Arbor, Michigan 48109} \\
\r {18} {\eightit Michigan State University, East Lansing, Michigan  48824} \\
\r {19} {\eightit University of New Mexico, Albuquerque, New Mexico 87131} \\
\r {20} {\eightit Osaka City University, Osaka 588, Japan} \\
\r {21} {\eightit Universita di Padova, Istituto Nazionale di Fisica 
          Nucleare, Sezione di Padova, I-35131 Padova, Italy} \\
\r {22} {\eightit University of Pennsylvania, Philadelphia, 
        Pennsylvania 19104} \\   
\r {23} {\eightit Istituto Nazionale di Fisica Nucleare, University and Scuola
               Normale Superiore of Pisa, I-56100 Pisa, Italy} \\
\r {24} {\eightit University of Pittsburgh, Pittsburgh, Pennsylvania 15260} \\
\r {25} {\eightit Purdue University, West Lafayette, Indiana 47907} \\
\r {26} {\eightit University of Rochester, Rochester, New York 14627} \\
\r {27} {\eightit Rockefeller University, New York, New York 10021} \\
\r {28} {\eightit Rutgers University, Piscataway, New Jersey 08854} \\
\r {29} {\eightit Academia Sinica, Taipei, Taiwan 11529, Republic of China} \\
\r {30} {\eightit Texas A\&M University, College Station, Texas 77843} \\
\r {31} {\eightit Texas Tech University, Lubbock, Texas 79409} \\
\r {32} {\eightit University of Tsukuba, Tsukuba, Ibaraki 305, Japan} \\
\r {33} {\eightit Tufts University, Medford, Massachusetts 02155} \\
\r {34} {\eightit University of Wisconsin, Madison, Wisconsin 53706} \\
\r {35} {\eightit Yale University, New Haven, Connecticut 06511} \\
\end{center}

\end{normalsize}

\begin{abstract}
The inclusive jet differential cross section has been measured
for jet transverse energies, $E_T$, from 15 to 440 GeV, in the
pseudorapidity region 0.1$\leq | \eta| \leq $0.7.
The results are based on 19.5 pb$^{-1}$
of data collected by the CDF collaboration at the Fermilab Tevatron collider.
The data are compared with QCD predictions for
various sets of parton distribution functions.
The cross section for jets with $E_T>200$\ GeV is significantly higher than 
current predictions based on 
O($\alpha_s^3$) perturbative QCD calculations.
Various possible explanations for the high-$E_T$\ excess are discussed.
\end{abstract}
\end{titlepage}
\label{sec-inclusive jet introduction}

        We present a precise measurement of the inclusive differential
 cross section for jet production in $p\bar{p}$ collisions at
 1.8 TeV. 
 Our measurement is compared to 
 next-to-leading order (NLO) perturbative QCD
 predictions~\cite{EKS}
 for jet transverse energies, $E_T$, from 15 to 440 GeV in the central
 pseudorapidity region 0.1$\leq | \eta |\leq$0.7, corresponding
 at highest $E_T$ to a distance scale 
 of O($10^{-17}$)~cm.

 The predictions depend on details of the parton distribution functions (PDFs) and on the
 strong coupling constant $\alpha_S$. Our measurement provides precise
 information about both
~\cite{{walter-alphas},{KUHLMANN}}.
 Apart from these 
 theoretical uncertainties, deviations of the predicted cross section
 from experiment
 could arise from physics beyond the Standard Model.
 In particular, the presence of quark substructure would enhance the 
 cross section at high $E_T$. 
 Previous measurements of inclusive jet production
  were performed with smaller data sets
 by CDF \cite{{CDF-Inclusive Jet-88},{CDF-Inclusive Jet-87}}
 and at lower energy by UA2 \cite{UA2} and CDF \cite{CDF-XT-analysis}.

The measurement described here is based on a data sample of 19.5 pb$^{-1}$ 
collected in 1992-93
with the CDF detector\cite{CDF-Detector} at the Tevatron collider.
The data were collected using several triggers
with jet $E_T$ thresholds of 100, 70, 50 and 20 GeV.
The 70, 50 and 20 GeV triggers were
prescaled by 6, 20 and 500, respectively.   
Cosmic rays and accelerator loss backgrounds were removed with cuts on 
event energy timing and on missing transverse energy, as described in
reference~\cite{CDF-Inclusive Jet-87}.
The remaining backgrounds are conservatively estimated to be $<$0.5\% in any
$E_T$ bin.

Jets were reconstructed using a cone algorithm\cite{CDF-Clustering} with 
radius
$R\equiv(\Delta\eta^2+\Delta\phi^2)^{1/2}=0.7$.
Here $\eta\equiv-$ln[tan$(\theta/2)$], where $\theta$ is the polar 
angle with respect to the beam line and $\phi$ is the azimuthal angle 
around the beam.
The QCD calculation
used a similar algorithm\cite{EKS}.
The ambient energy from
fragmentation of partons not associated with the 
hard scattering is subtracted.
No correction is applied for the energy falling outside the cone
because this effect is modelled by the NLO QCD calculations.

\begin{figure}
\centerline{\psfig{figure=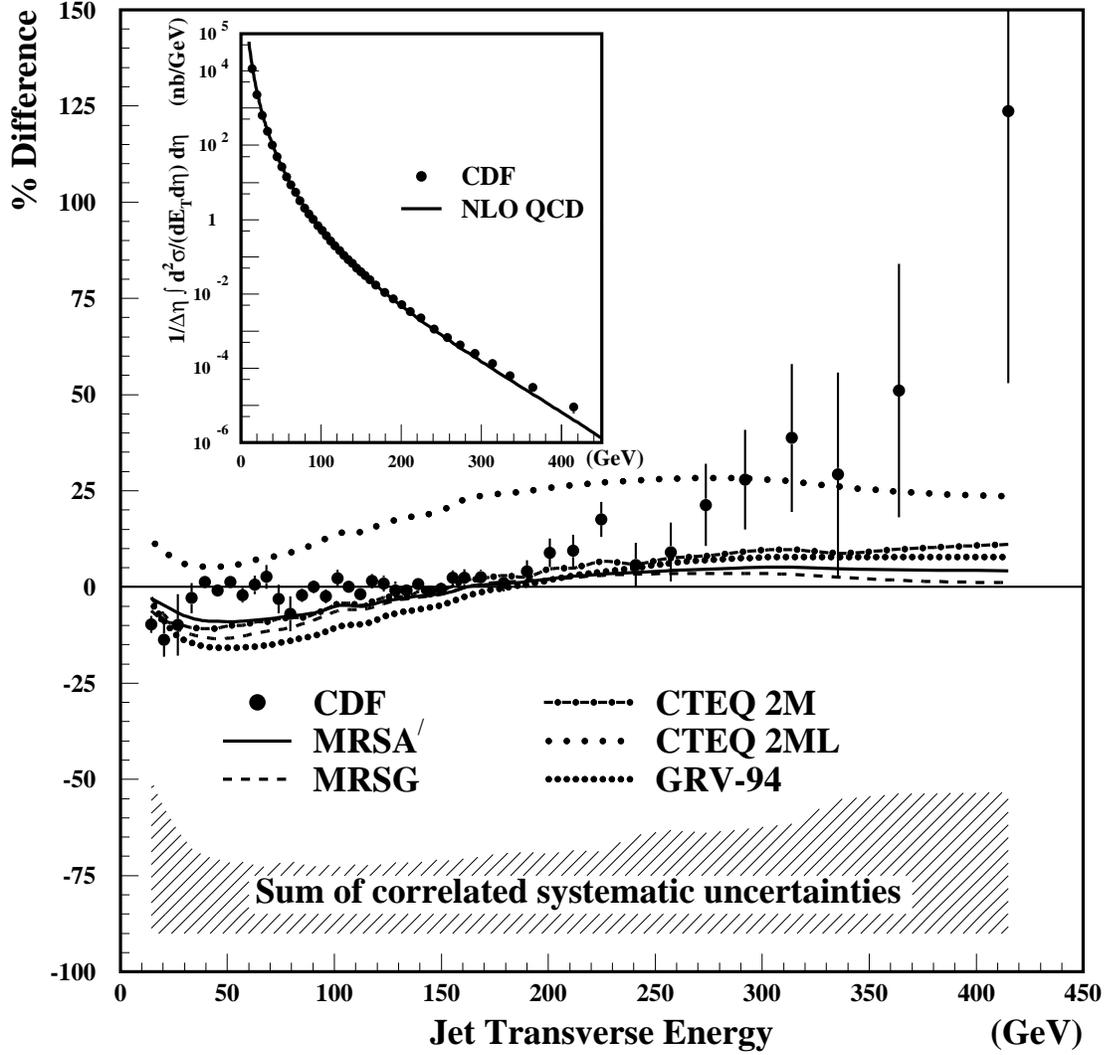,height=16cm,width=16cm}}
\caption{
The percent difference between the CDF inclusive jet cross section~(points) and
a next-to-leading order (NLO) QCD prediction using
MRSD0$^\prime$ PDFs.  The CDF data~(points) are compared directly 
to the NLO QCD prediction~(line) in the inset.
The normalization shown is absolute.
The hatched region at the bottom shows the quadratic sum of
correlated systematic uncertainties. NLO QCD predictions using
different PDFs are also compared with the one using MRSD0$^\prime$.}
\label{fig-qcd-qcd1}
\end{figure}

The measured jet $E_T$ spectrum is corrected for detector
and smearing effects caused by finite $E_T$ resolution
with the ``unsmearing procedure" described in \cite{CDF-XT-analysis}.
A Monte Carlo simulation, based on the ISAJET\cite{ISAJET} program and
 Feynman-Field\cite{FF} jet fragmentation tuned to
the CDF data, is used to determine detector response functions.
A trial true (unsmeared) spectrum is smeared with detector effects and 
compared to the raw data.  The parameters of the trial spectrum are iterated
to obtain the best match between the smeared trial spectrum and the
raw data.
We parameterize the unsmeared inclusive jet spectrum with
the functional form
\begin{eqnarray}
\label{eq:std_crv}
\frac{d\sigma(E_T^{True})}{dE_T^{True}} =
P_0\times (1-x_T)^{P_6} \times 10^{F(E_T^{True})} ,
\end{eqnarray}
where 
$F(x)= \sum_{i=1}^{5} P_{i}\times \left[\log(x)\right]^{i}$ with $E_T^{True}$
in GeV, $P_0 ... P_6$ are fitted parameters
and $x_T$ is defined as $2E_T^{True}/\sqrt{s}$.
The resulting fit of the smeared true spectrum to our data yields
$\chi^2$/degree-of-freedom $\equiv$ $\chi^2/d.f. = 29.9/34$.
The best-fit set of parameters for Eq.~\ref{eq:std_crv}, i.e.
the ``standard curve", are in Table\,\ref{table-fit param}.
Corrections to the measured $E_T$ and rate for each bin of the raw spectrum
are derived from the mapping of the standard curve to the smeared curve.
The corrected cross sections and statistical uncertainties are
in Fig.\,\ref{fig-qcd-qcd1} and 
in Table\,\ref{Table-Corrected Jet XSEC}.

\begin{figure}
\centerline{\psfig{figure=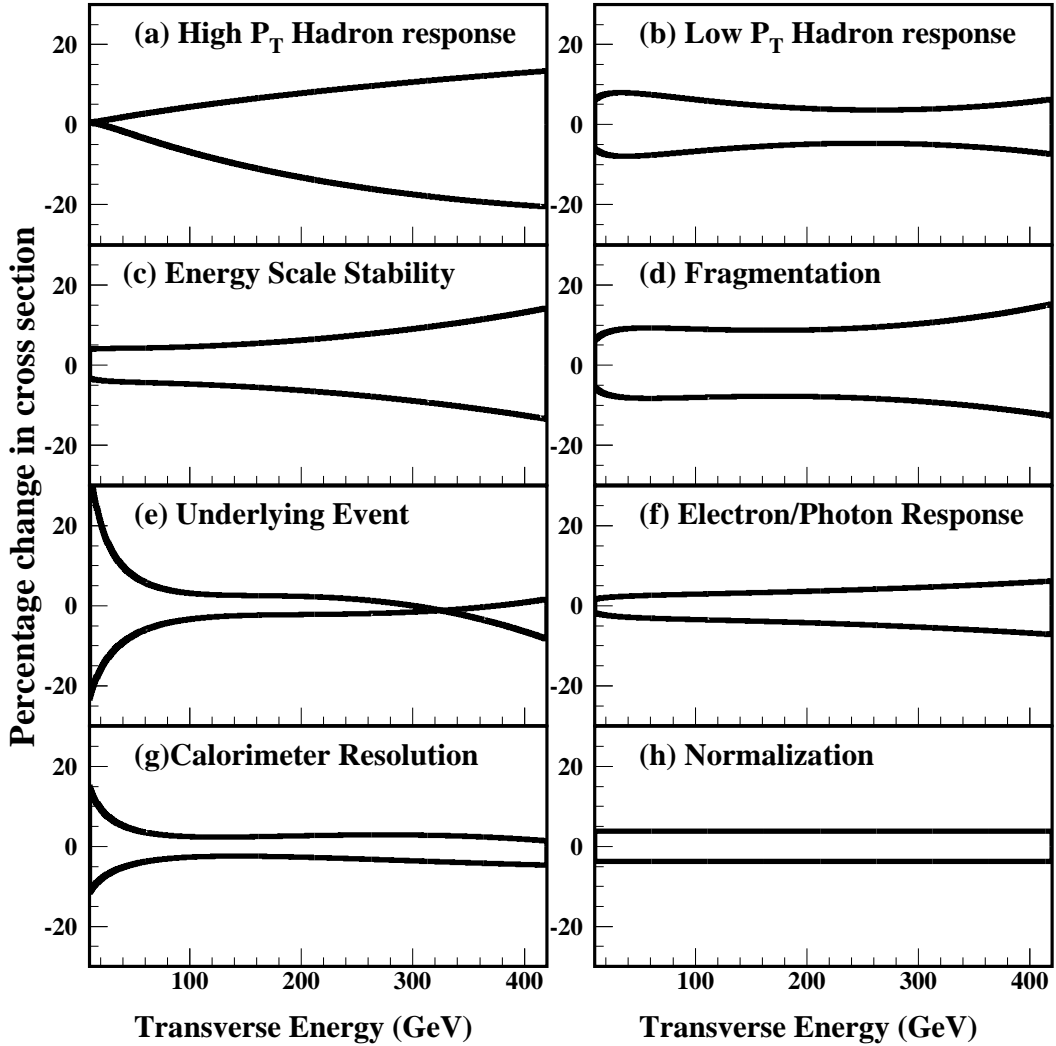,height=16cm,width=16cm}}
\caption{The percentage change in the inclusive jet cross section when
various sources of systematic uncertainty are changed by
$\pm$1-standard deviation from their nominal values.}
\label{Fig-sys-uncertainties}
\end{figure}

To evaluate systematic uncertainties, the procedure in reference
\cite{CDF-XT-analysis} is used.
New parameter sets for Eq.~\ref{eq:std_crv} are derived for 
$\pm 1$ standard deviation shifts in the unsmearing function 
for each source of systematic uncertainty.
The parameters for the eight largest systematic uncertainties are 
in Table\,\ref{table-fit param}.
They account for the following uncertainties:
(a) charged hadron response
at high $P_T$;
(b) the calorimeter response to low-$P_T$ hadrons;
(c) $\pm 1\%$ on the jet energy for the absolute calibration of the calorimeter;
(d)  jet fragmentation functions used in the simulation;
(e) $\pm 30\%$ on the underlying event energy in a jet cone;
(f) detector response to electrons and photons and
(g) modeling of the detector jet energy resolution.
An overall normalization uncertainty of $\pm$3.8\% was derived from the
uncertainty in the 
luminosity measurement ($\pm$3.5\%)
and the efficiency of the acceptance cuts ($\pm$1.5\%).
Additional tests of the unsmearing procedure, including
use of the HERWIG Monte Carlo program\cite{HERWIG}
to model jet fragmentation, 
were performed and the resulting
variations were found to be small.
Fig.~\ref{Fig-sys-uncertainties}(a--h) shows the percentage change
from the standard curve as a function of $E_T$ for each uncertainty.

In Fig.\,\ref{fig-qcd-qcd1} the corrected cross section is compared
with the NLO QCD prediction \cite{EKS} using
MRSD0$^\prime$ PDFs\cite{MRSD0},with
renormalization/factorization scale $\mu=E_T/2$.
These results show excellent agreement
in shape and in normalization for
$E_T\!<\!200$ GeV, while the cross section falls by
six orders of magnitude.
Above 200 GeV, the CDF cross section is significantly higher than
the NLO QCD prediction.
These data  are consistent with our previous measurement
\cite{CDF-Inclusive Jet-88}, which also shows an excess over
NLO QCD for the $E_T>280$ GeV region. A similar excess is observed
when we compare CDF data with HERWIG Monte
Carlo predictions.

The distributions of the physical variables in the 1192 events
above 200 GeV were examined carefully.  
Data distributions sensitive to the mismeasurement of jet $E_T$,
such as unbalanced jet $E_T$\ in dijet events,
show good agreement with detector simulation.
To look for time and luminosity dependent variations
(instantaneous luminosity increased with time),
the data were divided into seven time-ordered parts and analyzed
independently.
No significant time dependence was observed.
Finally, these events were individually scanned and no anomalies were
discovered.

No single experimental source of systematic uncertainty can account
for the high-$E_T$\ excess.
For example, in order to reconcile the measured CDF spectrum with NLO QCD
(MRSD0$^\prime$, $\mu=E_T/2$) predictions, we would
have to change the jet $E_T$ scale by an amount ranging
from 0.2\% at 175 GeV  to 5\% at
415 GeV, while keeping the change less than 0.1\% between 50 and 160 GeV.
No known feature of the detector, its calibration or the data
analysis permits such a change.
The effects of all 
possible combinations of the systematic uncertainties are included in the
comparison described below.

To analyze the significance of this excess we use four statistical tests:
signed and unsigned Kolmogorov-Smirnov~\cite{Roe},
Smirnov-Cram\`{e}r-VonMises~\cite{Roe}, and 
Anderson-Darling~\cite{{AD-stats1},{AD-stats2}}.
For this comparison we choose the MRSD0$^\prime$ PDFs which provide the
best description of our low $E_T$ data.
The eight sources of systematic uncertainty are treated individually
to include the $E_T$ dependence of each uncertainty.
The effect of finite binning and systematic uncertainties 
are modelled by a Monte Carlo calculation.  
The statistical tests over the full $E_T$ range 
are dominated by the higher precision data at 
low $E_T$; therefore, we test two ranges.
Between 40 and 150 GeV, the agreement between data and theory is $>$80\% 
for all four tests.  Above 150 GeV, however, each of the four methods yields 
a probability of 1\% that the excess is due to a fluctuation.

We have considered various sources of uncertainty in the theory.
The NLO QCD predictions have a weak dependence on the
renormalization/factorization scale $\mu$.
The change in $\mu$ scale from $2\,E_T$ to $E_T/4$ 
changes the normalization but maintains the shape for
$E_T>70$ GeV \cite{INCL-JET-PPBAR}.
For the NLO QCD calculations the renormalization and factorization
scales have been assumed to be equal. Varying these scales independently
also has little effect on the shape of the theoretical curve%
\cite{DSoper-mu-scale}.
However, soft gluon summation may lead to a
small increase in the cross section
at high $E_T$\cite{{Walter-pbarp},{Sterman-pbarp}}. In addition, 
the effect of higher order QCD corrections is not known.

The fractional difference between the MRSD0$^\prime$ \cite{MRSD0}
NLO QCD predictions  and predictions using different
choices of published PDFs, with $\mu=E_T/2$, is shown in
Fig.\,\ref{fig-qcd-qcd1}.
The excess of data over theory at high $E_T$ remains for
CTEQ2M\cite{CTEQ}, CTEQ2ML\cite{CTEQ}, GRV94\cite{GRV94},
MRSA$^\prime$\cite{MRSA} and MRSG\cite{MRSG} parton
distributions.
The variations in QCD predictions 
represent a survey of currently available distributions. 
They do not represent 
uncertainties associated with data used in deriving the PDFs.
Inclusion of our data in a global fit with those from other experiments
may yield a consistent set of PDFs that accommodate the high-$E_T$
excess within the scope of QCD~\cite{{KUHLMANN},{Stirling-fit}}.

The presence of quark substructure could appear as 
an enhancement of the
cross section at high $E_T$. This effect is conventionally parameterized in
terms of a contact term of unit strength between left-handed quarks,
characterized by the constant $\Lambda_C$ with units of
energy\,\cite{Eichten}.
While NLO standard model QCD predictions have been available for many years,
no calculation for compositeness at next-to-leading order 
[{\cal O($\alpha_s^3$)}] is available.
Therefore, we have compared our data to a LO QCD calculation including 
compositeness (using MRSD0$^\prime$) and have taken the approach 
of reference \cite{CDF-Inclusive Jet-88}.
We normalize the predicted cross section to the data 
over the $E_T$ range 95-145 GeV, where the effect of the contact term  with
$\Lambda_C\!>\!1000$ GeV is small.
The best agreement between this calculation and our data above 
$E_T\!>\!200$ GeV is for $\Lambda_C$ = 1600 GeV.
This hypothetical contact interaction is also expected to lead 
to dijet production with 
a more central angular distribution, and this analysis is underway.
However, until a realistic method for representing the theoretical
uncertainties from higher order QCD corrections and from 
the PDFs is found, any claim about the presence or absence
of new physics is not defensible.

In summary, we have measured the inclusive jet cross section in
the $E_T$ range 15-440 GeV
and find it to be in good agreement with NLO QCD predictions for
$E_T<200$ GeV using MRSD0$^\prime$ PDFs.
Above 200 GeV, the jet cross section is significantly higher
than the NLO predictions. 
The data over the full $E_T$ range are very precise.
They provide powerful constraints on QCD,  and demand 
a reevaluation of theoretical predictions and uncertainties 
within and beyond the Standard Model.

     We thank the Fermilab staff and the technical staffs of the
participating institutions for their vital contributions.  This work was
supported by the U.S. Department of Energy and National Science Foundation;
the Italian Istituto Nazionale di Fisica Nucleare; the Ministry of Education,
Science and Culture of Japan; the Natural Sciences and Engineering Research
Council of Canada; the National Science Council of the Republic of China; 
and the A. P. Sloan Foundation.


\begin{table}
\begin{center}
\caption{Parameters of the curves corresponding to $\pm$
1-standard deviation changes in the systematic uncertainties.}
\label{table-fit param}
\small
\vspace{0.2cm}  
\begin{tabular}{|l|c|c|c|c|c|c|c|  }
\hline
&
\multicolumn{1}{|c|}{$P_0$ (nb/GeV) } &
\multicolumn{1}{|c|}{$P_1$} &
\multicolumn{1}{|c|}{$P_2$} &
\multicolumn{1}{|c|}{$P_3$} &
\multicolumn{1}{|c|}{$P_4$} &
\multicolumn{1}{|c|}{$P_5$} &
\multicolumn{1}{|c|}{$P_6$} \\
\hline
Standard               &3.090$\times 10^{8}$ &$-$4.128&1.084  &$-$0.845  & 0.136  & 0.00279& 6.733  \\
High $P_T$ pion(+)     &3.000$\times 10^{8}$ &$-$4.110&1.083  &$-$0.847  & 0.135  & 0.00213& 6.500  \\
High $P_T$ pion(--)    &3.118$\times 10^{8}$ &$-$4.132&1.084  &$-$0.844  & 0.137  & 0.00299& 6.758  \\
Low $P_T$ pion (+)     &3.135$\times 10^{8}$ &$-$4.163&1.082  &$-$0.843  & 0.138  & 0.00342& 7.209  \\
Low $P_T$ pion (--)    &3.060$\times 10^{8}$ &$-$4.096&1.085  &$-$0.847  & 0.135  & 0.00216& 6.272  \\
1.0\% E. scale(+)      &3.174$\times 10^{8}$ &$-$4.122&1.083  &$-$0.846  & 0.136  & 0.00270& 6.434  \\
1.0\% E. scale(--)     &3.066$\times 10^{8}$ &$-$4.140&1.084  &$-$0.844  & 0.137  & 0.00274& 7.082  \\
Fragmentation  (+)     &3.152$\times 10^{8}$ &$-$4.161&1.082  &$-$0.843  & 0.138  & 0.00335& 7.214  \\
Fragmentation  (--)    &3.044$\times 10^{8}$ &$-$4.095&1.085  &$-$0.847  & 0.135  & 0.00220& 6.229  \\
Underly. Energy(+)     &6.630$\times 10^{8}$ &$-$4.314&1.067  &$-$0.840  & 0.141  & 0.00503& 8.045  \\
Underly. Energy(--)    &1.730$\times 10^{8}$ &$-$4.004&1.099  &$-$0.846  & 0.134  & 0.00122& 6.074  \\
Electron/$\gamma$(+)   &3.102$\times 10^{8}$ &$-$4.123&1.084  &$-$0.845  & 0.136  & 0.00271& 6.629  \\
Electron/$\gamma$(--)  &3.106$\times 10^{8}$ &$-$4.138&1.083  &$-$0.844  & 0.137  & 0.00287& 6.873  \\
Resolution     (+)     &2.422$\times 10^{8}$ &$-$4.082&1.090  &$-$0.845  & 0.136  & 0.00222& 6.645  \\
Resolution     (--)    &4.262$\times 10^{8}$ &$-$4.201&1.076  &$-$0.843  & 0.138  & 0.00366& 7.123  \\
\hline
\end{tabular}
\end{center}
\end{table}
\begin{table}
\begin{center}
\caption{The mean true jet $E_T$, cross section and
statistical uncertainty.}
\label{Table-Corrected Jet XSEC}
\vspace{0.2cm} 
\small
\begin{tabular}{|c|c|c|c|}
\hline
\multicolumn{1}{|c|}{$<\!E_T\!>$} &
\multicolumn{1}{|c|}{Cross Section} &
\multicolumn{1}{|c|}{$<\!E_T\!>$} &
\multicolumn{1}{|c|}{Cross Section} \\
\multicolumn{1}{|c|}{(GeV)} &
\multicolumn{1}{|c|}{(nb/GeV)} &
\multicolumn{1}{|c|}{(GeV)} &
\multicolumn{1}{|c|}{(nb/GeV)} \\
\hline
 14.5& ( 1.14$ \pm0.03 ) \times 10^4   $     & 133.8   & ( 8.50$ \pm0.12) \times 10^{-2}$ \\ 
 20.3& ( 2.31$ \pm0.12 ) \times 10^3   $     & 139.2   & ( 6.62$ \pm0.10) \times 10^{-2}$ \\ 
 26.9& ( 6.30$ \pm0.56 ) \times 10^2   $     & 144.5   & ( 5.00$ \pm0.08) \times 10^{-2}$ \\ 
 33.3& ( 2.36$ \pm0.09 ) \times 10^2   $     & 149.9   & ( 3.92$ \pm0.07) \times 10^{-2}$ \\ 
 39.5& ( 1.02$ \pm0.01 ) \times 10^2   $     & 155.3   & ( 3.13$ \pm0.06) \times 10^{-2}$ \\ 
 45.5& ( 4.89$ \pm0.06 ) \times 10^1   $     & 160.7   & ( 2.46$ \pm0.05) \times 10^{-2}$ \\ 
 51.3& ( 2.61$ \pm0.04 ) \times 10^1   $     & 168.4   & ( 1.75$ \pm0.03) \times 10^{-2}$ \\ 
 57.0& ( 1.42$ \pm0.03 ) \times 10^1   $     & 179.2   & ( 1.10$ \pm0.02) \times 10^{-2}$ \\ 
 62.7& ( 8.62$ \pm0.21 ) \times 10^0   $     & 189.0   & ( 7.34$ \pm0.20) \times 10^{-3}$ \\ 
 68.3& ( 5.43$ \pm0.16 ) \times 10^0   $     & 200.7   & ( 5.11$ \pm0.17) \times 10^{-3}$ \\ 
 73.9& ( 3.24$ \pm0.13 ) \times 10^0   $     & 211.5   & ( 3.41$ \pm0.13) \times 10^{-3}$ \\ 
 79.4& ( 2.05$ \pm0.10 ) \times 10^0   $     & 224.6   & ( 2.25$ \pm0.09) \times 10^{-3}$ \\ 
 85.0& ( 1.44$ \pm0.02 ) \times 10^0   $     & 240.9   & ( 1.14$ \pm0.06) \times 10^{-3}$ \\ 
 90.5& ( 1.02$ \pm0.02 ) \times 10^0   $     & 257.2   & ( 6.67$ \pm0.47) \times 10^{-4}$ \\ 
 95.9& ( 6.94$ \pm0.13 ) \times 10^{-1}$     & 273.5   & ( 4.31$ \pm0.38) \times 10^{-4}$ \\ 
101.4& ( 5.18$ \pm0.11 ) \times 10^{-1}$     & 292.0   & ( 2.50$ \pm0.25) \times 10^{-4}$ \\ 
106.8& ( 3.64$ \pm0.05 ) \times 10^{-1}$     & 313.7   & ( 1.35$ \pm0.19) \times 10^{-4}$ \\ 
112.2& ( 2.64$ \pm0.04 ) \times 10^{-1}$     & 335.3   & ( 6.37$ \pm1.30) \times 10^{-5}$ \\ 
117.6& ( 2.00$ \pm0.04 ) \times 10^{-1}$     & 364.0   & ( 3.03$ \pm0.66) \times 10^{-5}$ \\ 
123.0& ( 1.48$ \pm0.03 ) \times 10^{-1}$     & 414.9   & ( 9.05$ \pm2.86) \times 10^{-6}$ \\ 
128.4& ( 1.10$ \pm0.03 ) \times 10^{-1}$&         &                             \\ 
\hline
\end{tabular}
\end{center}
\end{table}
\end{document}